\documentclass[aps,prd,onecolumn,preprintnumbers,superscriptaddress,nofootinbib,floatfix]{revtex4-2}
\usepackage[utf8]{inputenc}
\usepackage{amssymb}
\usepackage{tensor}
\usepackage{xspace} 
\usepackage[pdftex]{hyperref}
\usepackage{hyperref}
\usepackage{graphicx}
\usepackage[caption=false]{subfig}
\usepackage{breakcites}
\usepackage{ragged2e}

\usepackage{setspace}
\hypersetup{
colorlinks,linkcolor=blue,citecolor=blue,urlcolor=blue
}

\usepackage{natbib}
\usepackage{tablefootnote}
\usepackage{footnote}
\usepackage{amsmath}
\usepackage{amssymb}
\usepackage[T1]{fontenc}

\begin{document}

\title{Energy Extraction via Magnetic Reconnection in Lorentz breaking Kerr-Sen and Kiselev Black Holes }

\newcommand{\UNISA}{\affiliation{Dipartimento di Fisica ``E.R Caianiello'', Università degli Studi di Salerno,\\ Via Giovanni Paolo II, 132 - 84084 Fisciano (SA), Italy}}
\newcommand{\INFN}{\affiliation{Istituto Nazionale di Fisica Nucleare - Gruppo Collegato di Salerno,\\ Via Giovanni Paolo II, 132 - 84084 Fisciano (SA), Italy.}}

\author{Amodio Carleo}
\email{acarleo@unisa.it}
\UNISA\INFN

\author{Gaetano Lambiase}
\email{glambiase@unisa.it}
\UNISA\INFN

\author{Leonardo Mastrototaro}
\email{lmastrototaro@unisa.it}
\UNISA\INFN

\date{\today}
%\setlenght\parindent{}

%\begin{frontmatter}

\begin{abstract}
  Black holes can accumulate a large amount of energy, responsible for highly energetic astrophysical phenomena 
  Recently, fast magnetic reconnection (MR) of the magnetic field was proposed as a new way to extract energy and in this paper, we investigate this phenomena in a bumblebee Kerr-Sen BH. We find that the presence of the charge parameter strongly changes the simple Kerr case, making this extraction mechanism possible even for not extremely rotating black holes ($a \sim 0.7$). We also show that, under appropriate circumstances, MR is more efficient compared to the Blandford-Znajek mechanism. We finally compare these results with quintessence black-hole solutions not finding and enhancement respect to Kerr solution.    
\end{abstract}

%\begin{keyword} Energy Extraction \sep%
%    Magnetic Reconnection \sep Lorentz Symmetry  \sep Kerr-Sen \sec dark energy \sec quintessence 
%\end{keyword}

%\keywords{NGC 4490; ULX; Wolf-Rayet; Mergers }

\maketitle

\section{Introduction}

Despite the no-hair theorem, according to which a black hole can not have its own magnetic field, typically astrophysical black holes are immersed in an external magnetic field, generated from the accretion disk. The recent analysis, published by the EHT Collaborations~\cite{M87} on polarized emission around the supermassive BH in the center of M87*, has confirmed the existence of a magnetic field around the BH. It is well known that energy extraction from a spinning black hole can in principle explain some of the most energetic astrophysical events in the Universe, like relativistic jets from active galactic nuclei (AGN) and gamma-ray bursts (GRBs). The high energy released in such events is believed to be a fraction of the potential energy of the matter falling towards the black hole. In the last years, however, the hypothesis that the energy of the black hole itself can be stolen is gaining ground as a complementary mechanism.

The first process studied of this type is the Penrose one: the existence of a space-like Killing vector inside the ergosphere allows the existence of negative energy states in this region. Since they exist outside the horizon, an infalling particle into the black hole carries negative energy and angular momentum. Hence, the total energy of the black hole decreases. From considerations of variations of energy $\delta E$, momentum $\delta J$ and mass $\delta M$ of the black hole, one finds \cite{Penrose:1969pc,Penrose:1971uk} that $\delta J \leq \delta M/\Omega_{H}$, where $\Omega_{H}$ is the angular velocity of the horizon. Since $J$ can never become negative, we can define an irreducible mass of a Kerr black hole for unit mass as \cite{Christodoulou:1970wf}
\begin{equation} \label{eq1}
M^{2}_{irr} = \dfrac{1}{2}\Big(  1 + \sqrt{1-a^{2}} \Big) \,\ ,
\end{equation}
where $a= J~M$ is the angular momentum parameter. However, the Penrose process is not an efficient way of extracting energy from a black hole \cite{Wald:1974kya}: it requires that the relative velocity of the two created particles is greater than half of the speed of light $c$  \cite{Wald:1974kya} and the expected rate of such events is believed to be very rare. Hence, other ways should be investigated.

In principle, any such process can extract a maximum energy equal to $E_{max}^{rot} = M - M_{irr}$, which means  a maximum efficiency of $\eta_{rot} = \frac{M-M_{irr}}{M} \simeq 29 \% $  for $a=1$. Despite this maximum efficiency, the amount of energy strongly depends on the angular parameter $a$, halving as soon as $a$ passes from 1 to 0.9. Nowadays, the best candidate for energizing a wide range of highly energetic astrophysical phenomena is the Blandford-Znajek (BZ) process (cfr. \cite{BZ1,BZ2,BZ3,BZ4}). When a zero-charge black hole spins in a uniform magnetic field $B$ generated by the accretion disk, the spin induces an electric field with $E \cdot B \neq 0 $ . This induced electric field captures any external charged particle, and, as $E \cdot B$ has opposite signs at the poles and the equator \cite{King:1975tt}, carries them either to the poles or to the equator. If the black hole is surrounded by a charged medium, a real charged current through the black hole is formed \cite{BZ-1977ds} and twists the magnetic field lines into a tight helix, draining energy from the poles and powering the so-called jets.  It has been shown that the strongly magnetized BZ mechanism is more effective in General Relatvity than the non-magnetized neutrino annihilation processes to power gamma ray-burst (GRB) jets for the same BH spin parameter and accretion rate (see e.g. \cite{Liu:2017rwh}). The picture is different in theories beyond general relativity as shown in Ref.~\cite{Lambiase:2020iul,Lambiase:2020pkc,Lambiase:2022ywp}

Beyond the more exotic Hawking radiation, another extraction mechanism has recently been proposed. In Ref.~\cite{Comisso}, the authors showed that, when a Kerr BH is immersed in an externally supplied magnetic field, reconnection of magnetic field lines within the ergosphere can generate negative energy particles (relative to infinity) that fall into the event horizon and positive energy particles which steal energy from the black hole. In other words, Magnetic reconnection accelerates part of the plasma in the direction of the black hole rotation and another part in the opposite direction which falls into the black hole. In particular, the frame-dragging of the spinning black hole generates antiparallel magnetic field lines just above and just below the equatorial plane. The change of the magnetic field at the equatorial plane produces an equatorial current sheet interrupted by the formation of plasmoids, which drives fast magnetic reconnection. This rapidly converts the available magnetic energy into plasma particle energy. Comisso and Asenjo \cite{Comisso} analytically found that this channel is several times more efficient than the BZ one, but energy extraction is possible only for an extreme rotating black hole, $a \sim 1$, and in presence of strongly magnetized plasma, $\sigma > 1/3$, where $\sigma$ is the plasma magnetization.  In this scenario, the maximum power extracted is when the dominant reconnection point is close to the event horizon and corresponds to $P_{extr}^{max} \sim 0.1 M^{2}\sqrt{\sigma} w_{0}$, where $M$ is the black hole mass, $w_{0}$ the plasma enthalpy density and we are considering a collisionless plasma regime\footnote{In the case of a collisional plasma, $P_{extr}^{max}$ is one order of magnitude greater}. The minimum $\sigma$ value for extracting energy is $\sim 1/3$ but the efficiency of such a process is greater than 1 only for $\sigma \gg 1$. Similar values are assumed in supermassive black holes in active galactic nuclei (AGNs), where typically $\sigma \sim 10^{4}$ or larger, as in our galactic center \cite{sigma2010}. Indeed, magnetic reconnection close to the black hole is often conjectured to induce X-ray and near-infrared flares \cite{sigma2010}. An important signature of this new energy extraction way is its transient nature in opposition to the continuous nature of the BZ process. The reason for this bursty behaviour is the time it takes to accumulate magnetic energy which requires appropriate dynamics of the magnetic field lines configuration. This feature reinforces the idea of its role in relativistic jets. Furthermore magnetic reconnection seems several times more energetic respect to BZ process: the power ratio $P_{MR}/P_{BZ}$ is largely greater than 1 for an extended range of plasma magnetization $\sigma$. On the other hand, energy extraction via fast magnetic reconnection, while increasing for higher magnetization values,  is always subdominant to the Blandford-Znajek  for $\sigma \rightarrow  \infty$ . Differently from the BZ process, in which the extraction is merely electromagnetic, the magnetic reconnection requires non-zero particles inertia. Furthermore, it differs from the Penrose mechanism since a magnetic field is not required in the latter case. The common point is the existence of an ergosphere and a dragging phenomenon of space-time around the black hole. This implies that any static solution must be discarded and, indeed, high spin values are favoured.

In the wake of Ref.~\cite{Comisso}, the role of a Lorentz parameter $l$ in energy extraction has recently been investigated Ref.~\cite{mohsen}. The author found that energy extraction power from a rotating BH solution with broken fundamental Lorentz symmetry is, in some cases, more efficient than in the classical Kerr solution. Lorentz symmetry, although fundamental, may be violated in high energy limit \cite{lorentz}. In \cite{mohsen}, the bumblebee gravity model was adopted \cite{kerr-like-bumblebee}: the violation of the Lorentz symmetry is due to a non-zero vacuum expectation value (VEV) of the so-called bumblebee vector field coupled to the curvature of the space-time. In recent years, this type of solution has been carefully considered, due to the presence of additional parameters which provide a richer phenomenology framework.

Moreover, typically, BZ and magnetic reconnection mechanisms do not contemplate any electric charge on the black hole, since the electric charge is usually neglected. However, as already noted by Wald in \cite{Wald:1974kya}, if the medium surrounding the black hole contains mobile charges, then a spinning hole quickly acquires an electric charge, whose effect is to nullify the electric field which drives the BZ mechanism \cite{King2021}. This has a double implication: either the jet engine we observe in astrophysics is not the central black hole, or, in cases where there is indeed mobile plasma around, the BZ mechanism is less effective than we expect. Furthermore, an electric charge $Q=\sqrt{b M}$ ($b$ is the charge parameter) strongly affect the dynamics around the BH, especially shifting \cite{charge2019} significantly the innermost stable circular orbit (ISCO), whose radius is one of the most notable parameters for magnetic reconnection energy extraction. Hence the importance of also considering a charge, albeit a small one  \footnote{High value of charge has been associated with severe instability ( cfr. Ref.~\cite{instabilities})}.

In this paper, based on the previous considerations, we investigate energy extraction via MR for a Kerr-Sen-like black hole in the Einstein-bumblebee theory of gravity, i.e. a Kerr-Sen black hole with the addition of the violation of Lorentz symmetry. In particular, we discuss the effect of the Lorentz and charge parameters on this new promising energy extraction channel. The outline of this paper is as follows. In Sec. \ref{sec2}, we briefly present a Kerr-Sen-like solution with broken Lorentz symmetry given by a bumblebee field and derive some relevant quantities. In Sec. \ref{sec3} , energy extraction is discussed after selecting the most promising values for $b$ and $\ell$. In Sec. \ref{sec4} we compute the power ratio between the magnetic reconnection channel and BZ one. In Sec. \ref{sec5} we analyze the parametric space for different models to find bounds on the parameters.In Sec. \ref{sec6} we adopt the same strategy to the quintessence model. We also consider the case of dust and radiation. Finally, conclusions are presented in the last section \ref{sec7}. Unless otherwise stated, we will use $G=c=1$.

\section{Kerr-Sen-like black hole in bumblebee gravity}\label{sec2}
A proper Kerr-Sen solution derives from the heterotic string theory, whose metric solutions, $G_{\mu \nu}$, are related to the Einstein metric by $g_{\mu \nu}= e^{-\Phi}G_{\mu \nu}$, where $\Phi$ is the dilaton field.  A Kerr-Sen-like solution, however, can be pulled out of a generalized form of radiating stationery axially symmetric black-hole metric, once an appropriate background has been chosen that extends the Hilbert-Einstein action. Heterotic string theory is one of the primary candidates to describe quantum gravity, with some relevant differences from GR, which make it visible in crucial phenomenological aspects, such as the shadow of a black hole (see e.g. \cite{Xavier_2020} and \cite{TestKerrSen}). Although it has been shown that a Kerr-Sen black hole has a larger shadow than its general relativity analogue (Kerr-Neumann BH), this effect, already really small, is negligible when the charge is low. \\
Since we are interested in a  background which is  Lorentz violating, one possibility is to consider the action of Einstein-Bumblebee gravity \cite{Jha:2020}, namely 
\begin{equation} \label{eq2}
    S=\int d^{4} x \sqrt{-g}\left[\frac{1}{16 \pi G_{N}}\left(\mathcal{R}+\varrho B^{\mu} B^{\nu} \mathcal{R}_{\mu \nu}\right)-\frac{1}{4} B^{\mu \nu} B_{\mu \nu}-V\left(B^{\mu}\right)\right]
\end{equation}

where $\varrho$ is a non-minimal coupling constant between gravity and the bumblebee vector field $B^{\mu}$, whose potential is indicated with $V$.  The resulting effective field theory is Lorentz breaking due to non-zero vacuum expectation value (VEV) of the field $B^{\mu}$, i.e. $\langle B^{\mu}\rangle = Z^{\mu}$,  provided that the potential $V$ have a minimum by the condition $B_{\mu}B^{\mu} \pm Z^{2} = 0$, with $Z^{2}$ is a real positive constant. The generalized form of radiating stationary axially symmetric black-hole metric in  Boyer-Lindquist coordinate can be written down as
\begin{equation} \label{ds}
\begin{array}{l}
d s^{2}=-\gamma(\zeta, \theta) d t^{2}+a[p(\zeta)-q(\theta)]\left(d \zeta^{2}+d \theta^{2}\right) \\
+\quad\left\{[1-\gamma(\zeta, \theta)] q^{2}(\theta)+p(\zeta) q(\theta)\right\} d \phi^{2}-2 q(\theta)[1-q(\zeta, \theta)] d t d \theta
\end{array}
\end{equation}

where $a$ is, at this level, just a dimensional constant. Assuming a space-like bumblebee field, which acquires a pure radial VEV, and naming $\ell=\varrho b_{\mu}b^{\mu} $, we arrive at the rotating metric in the bumblebee gravity

\begin{equation}
    d s^{2}=-\left(1-\frac{2 M r}{\rho^{2}}\right) d t^{2}-\frac{4 M r a \sqrt{1+\ell} \sin ^{2} \theta}{\rho^{2}} d t d \varphi+\frac{\rho^{2}}{\Delta} d r^{2}+\rho^{2} d \theta^{2}+\frac{A \sin ^{2} \theta}{\rho^{2}} d \varphi^{2}
    \label{metric}
\end{equation}
where
\begin{eqnarray}
\Delta&=&\frac{r(r+b)-2 M r}{1+\ell}+a^{2} \,\ ,\\
A &=& \left[r(r+b)+(1+\ell) a^{2}\right]^{2}-\Delta(1+\ell)^{2} a^{2} \sin ^{2} \theta \,\ , \\
\rho^{2}&=&r(r+b)+(1+\ell) a^{2} \cos ^{2} \theta \,\ .
\end{eqnarray}
The parameter $a$ has the role of angular momentum, $J=a/M$, while $b$ is the charge parameter, $Q=\sqrt{b M}$. Clearly, if $\ell \rightarrow 0$, (3) becomes the usual Kerr-Sen metric \cite{Sen:1992ua}; for $\ell \rightarrow 0$ and $b \rightarrow 0$ we recover classical Kerr solution; for the parameter $b = 0$ it turns into Kerr-like
metric and for both $b = 0$ and $a = 0$ the metric lands onto Schwarzschild-like metric. Since we are interested in magnetic reconnection in the ergosphere,  from the conditions $g_{00} = 0$ and $g_{rr} \rightarrow \infty$, we obtain  its inner and outer radius:
\begin{equation} \label{rin}
   r_{in} = M-\frac{b}{2} + \frac{\sqrt{(b-2 M)^{2}-4 \tilde{a}^{2}}}{2}, \; \; r_{out} = M-\frac{b}{2} + \frac{\sqrt{(b-2 M)^{2}-4 \tilde{a}^{2} \cos^{2} \theta}}{2} \end{equation}
where $\tilde{a} \doteq a \sqrt{1+\ell}$. They corresponds to outer event horizon and static limit, respectively. It is clear that an event horizon exists if and only if  $\ell>-1$, in addition to

\begin{equation*}
    l \leq \Big( \dfrac{b-2}{2a} \Big)^{2} -1,
\end{equation*}
which is the condition to have a real $r_{in}$ ($r_{out}$ is always positive at the equator).\\

In Ref.~\cite{Xavier_2020}, the shadow of a Kerr-Sen black hole was compared with that of Kerr-Newmann, finding that the first always has a larger shadow, for the same physical parameters and observation conditions. Adding the Lorentz parameter further changes the shadow: it gets shifted (w.r.t. the ideal center) towards the right for positive $l$ and towards the left for negative $l$ when $a$ and $b$ are fixed. In particular, fitting M87* as a Kerr-Sen-like black hole, an upper limit to $l$ has recently been found to be $\ell < 0.63$ \cite{Jha_2022}. Hence, we use the range $\ell \in (-1, 0.6] $ in the following, while for the charge, we take values as low as possible, i.e. $b/M \leq 0.3 $ \footnote{Since spinning black holes with electric charge have an intrinsic magnetic field, for high $b$ values one should also consider how the magnetic field of the disk changes, from whose orientation energy extraction depends.}. Besides, from the expression of $r_{in}$, an upper limit on $a$ can be set, i.e. $a \leq (2-b)/(2\sqrt{1+l})$, which is maximum at $\ell \lesssim -1$ and minimum for $\ell =0.6$, being $b$ fixed. There is not a similar situation in Kerr.   \\
We consider the equation of motion along the worldline of a particle in the space-time~\ref{metric}: 
\begin{equation} \label{L}
  \mathcal{L} =  \dfrac{1}{2}g_{\nu \mu} \dfrac{dx^{\mu}}{d\lambda}\dfrac{dx^{\nu}}{d\lambda} = \epsilon \,\ ,
\end{equation}
where $\lambda$ is the affine parameter and $\epsilon = -1/2$ or $\epsilon = 0$ depending on whether the trajectory is time-like or null-like respectively. Since we focus on circular orbits in the plane of the equator, we set $\theta = \pi/2$, getting an equation for $r$
\begin{equation} \label{eq7}
    \Dot{r}^2 = \dfrac{\Delta}{\rho^2} \Big[ 2\epsilon - \Big(  \dfrac{2Mr-\rho^2}{\rho^2} \Big)\Dot{t}^2  - \dfrac{A}{\rho^2} \Dot{\phi}^2 + \dfrac{4Mra\sqrt{1+l}}{\rho^2} \Dot{t}\Dot{\phi} \Big] \,\ ,
\end{equation}
where the dot means derivative with respect to the affine parameter $\lambda$. The right-hand member plays the role of an effective potential, hence in the following we call it $V_{eff}$. From the definition of energy, $E= -\partial \mathcal{L}/\partial \Dot{t}$, and angular momentum, $L= \partial \mathcal{L}/ \partial \Dot{\phi}$, one finds the  rather general relations:

\begin{equation} \label{E}
   E = -g_{tt}\Dot{t} - g_{t\phi}\Dot{\phi}, \; \; L = g_{\phi \phi} \Dot{\phi} + g_{t\phi} \Dot{t} \,\ ,
\end{equation}
from which 
\begin{eqnarray}
\dot{t}&=&-\frac{E}{g_{tt}}\left[ 1 + g_{t\phi} \left(S + \frac{g_{t\phi} }{g_{tt}}\right) \left(g_{\phi\phi}- \frac{g_{t\phi}^{2}}{g_{tt}}\right)^{-1}\right] \\
\dot{\phi}&=&E \left(S + \frac{g_{t\phi}}{g_{tt}}  \right) \left(g_{\phi\phi}- \frac{g_{t\phi}^{2}}{g_{tt}}\right)^{-1} \,\ ,
\end{eqnarray}
where $S \doteq L/E$ is the impact parameter.

\begin{figure*}[t!!!]
\centering
\vspace{0.cm}
\includegraphics[width=0.46\columnwidth]{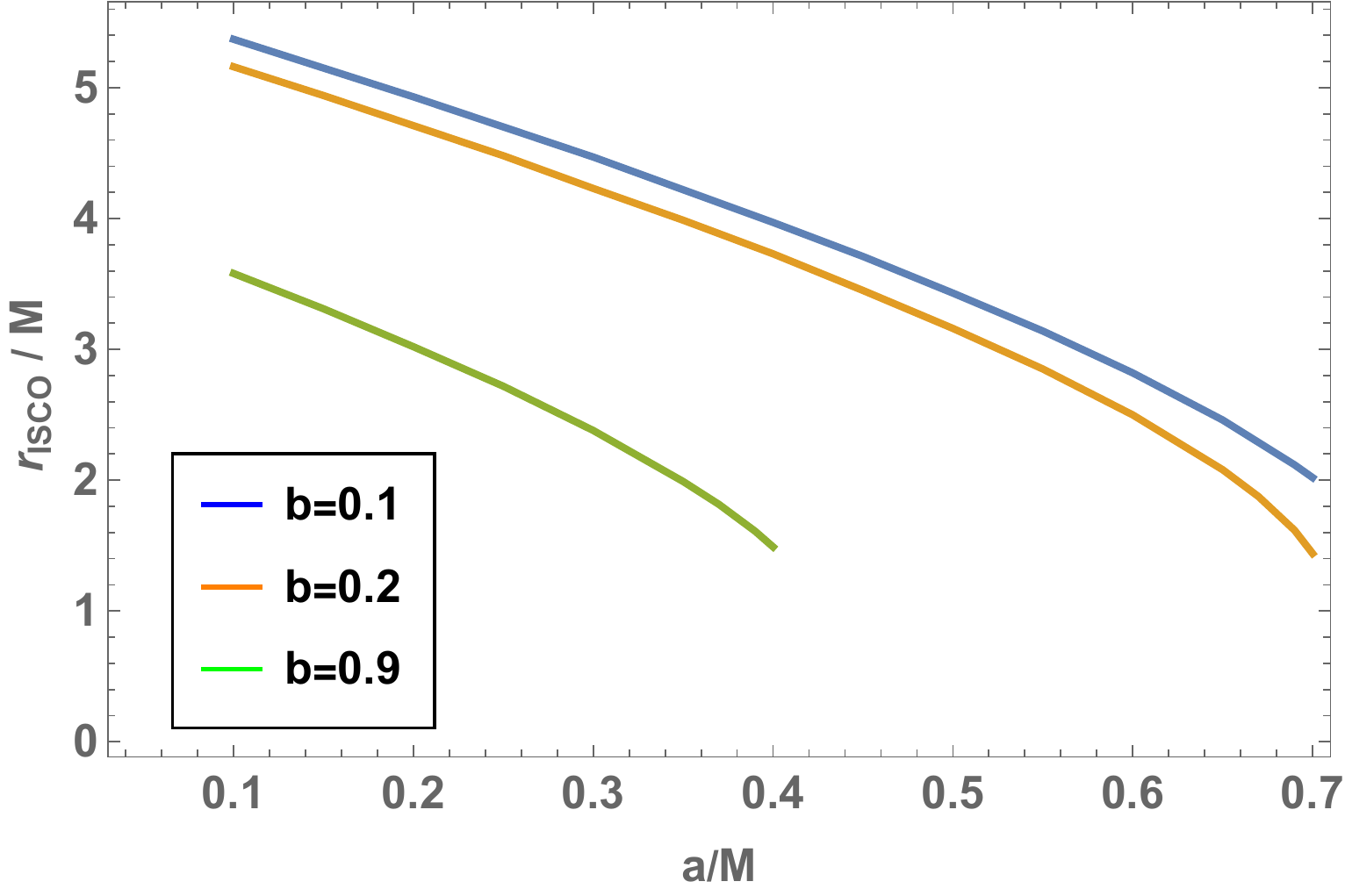}
\hspace{.35
cm}
\includegraphics[width=0.485\columnwidth]{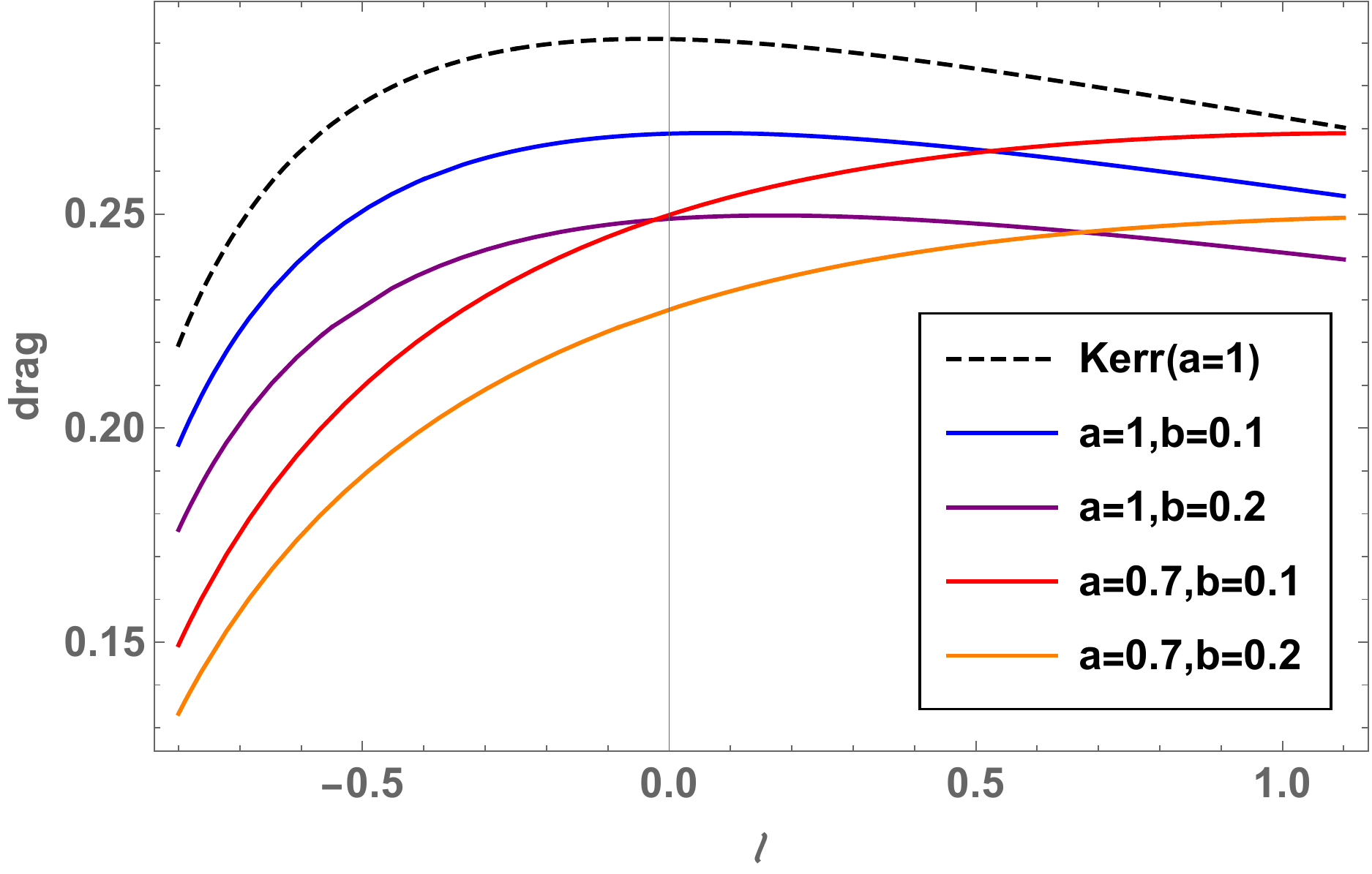}

\caption{(a) Trend of $r_{isco}$ in Lorentz violating Kerr-Sen black hole for different values of the charge parameter $b$ and with $l=0.6$ fixed. Note that not all values of the spin parameter $a$ are always available. (b) Trend of the drag effect, $\Omega=\dot{\phi}/\dot{t}$, as function of the parameter $\ell$, for different models, with the simplification $L=0$ and $r=1.5$, which is approximately in the middle ergosphere. Notice the supremacy of the Kerr case.}
\label{fig:1}
\end{figure*}

To find photonsphere and innermost stable circular orbits, $r_{ph}$ and $r_{isco}$ respectively,  we impose the circularity condition $\dot{r}=0$. For $r_{ph}$, we need only two simultaneous conditions from Eq.~(\ref{eq7}): $V_{eff}(r)=0$ and $V_{eff}'(r) = 0$, where prime stands for radial derivative. For $r_{isco}$, we need to impose also $V_{eff}''(r) = 0$. Unlike from Kerr solution, here it is impossible to obtain analytical expressions for $r_{ph}$ and $r_{isco}$, and numerical analysis is the only way. The effect of charge, with the same $l$ value, is to restrict the $r_{isco}$ (Fig. \ref{fig:1}(a)), and this is a priori a favorable condition for having magnetic reconnection in the ergosphere.

Increasing $b$ means also having  smaller $r_{out}$, i.e. a smaller ergoregion, whose extension is linked with the amount of energy expelled (see next section). However, this disadvantage vanishes when $a<1$, where the presence of $\ell \not= 0$ and $b \not= 0$, allows for larger ergoregions with respect to Kerr case. Furthermore, $r_{isco}$ decreases as a function of $b$ much faster than $r_{out}$ increases as $b$ increases. On the other hand, frame dragging, $\Omega=d\phi/dt$, is reduced as spinning decreases. However, while Kerr remains the best case,  in the bumblebee-Kerr-Sen models \footnote{In the following, we often indicate with \textit{model} a chosen  set of the parameters $(a,b,l)$.}, the decrease is negligible if the values of $\ell$ are chosen appropriately. As shown in Fig. \ref{fig:1} (b), when $a \rightarrow 1$, the maximum dragging value, $\Omega_{max}$, is obtained for negative $l$, while for $a<1$ positive values are favored. In particular, larger $b$ values are associated with reduced $\Omega_{max}$. Using all this information, we find that the small spin value for which $r_{isco} < r_{out}$ is $a=0.7$ for $b=0.2$ and $l=0.6$. For all other possible  values of the parameters $b$ and $\ell$, this condition is realized only at higher spin values. We also found that, given a value of the charge $b$ (with $b<0.4$), the smallest value of $\ell$ in order to have $r_{isco}<r_{out}$ is $\ell=-b$. \\

Following Ref.~\cite{Comisso}, we assume that magnetic reconnection happens in the the bulk plasma which stably rotates around the black hole. Since the orbit is supposed circular, the angular velocity is keplerian, $w_{K}=\Dot{\phi}/\Dot{t}$. From the r-component of the Euler-Lagrange equation and with $g_{r\mu} = 0$ if $\mu \not= r$  and $\dot{r}=\ddot{r} = 0$, one gets an equation for $w_{K}$:
\begin{equation*}
    g_{\phi \phi , r} w_{K}^{2} + 2 g_{t \phi, r} w_{K} + g_{t t, r} = 0 \,\ ,
\end{equation*}
whose double solution is:
\begin{equation} \label{eq11}
  w_{K}=\frac{-2 \tilde{a} M \pm (b+r) \sqrt{2 M (b+2 r)}}{b^{3}-2 \tilde{a}^{2}M +4 b^{2} r+5 b r^{2}+2 r^{3}} \,\ .
  \end{equation}
The upper sign refers to co-rotating orbits (i.e. $L>0$), while the lower sign applies to counter-rotating orbits.  In order to stay inside the ergosphere, plus sign is the only choice. When $a,b,\ell \rightarrow 0$,   Eq.~(\ref{eq11}) becomes $w_{K} = \pm \sqrt{M/r^3}$ which is keplerian angular velocity for Schwarzchild black hole.\\

Before analyzing the energy extraction, it is necessary to calculate some quantities in a locally non rotating (Minkowskian) \textit{zero-angular-momentum-observer} (ZAMO) frame \cite{ZAMO}, whose square line element is given by $d s^{2}=-d \hat{t}^{2}+\sum_{i=1}^{3}\left(d \hat{x}^{i}\right)^{2}=\eta_{\mu \nu} d \hat{x}^{\mu} d \hat{x}^{\nu}$, where $d \hat{t}=\alpha d t, \quad d \hat{x}^{i}=\sqrt{g_{i i}} d x^{i}-\alpha \beta^{i} d t$ with $i=1,2,3$ and no summation over $i$. Quantities in the ZAMO frame are denoted with hats. Here, $\alpha$ and $\beta{i}$ are the lapse function and the shift vector $(0,0,\beta^{\phi})$, respectively, i.e.
\begin{equation} \label{alpha}
    \alpha = \Big( 1- \dfrac{2Mr}{\rho^{2}} + \dfrac{4M^{2}r^{2}\tilde{a}^{2}}{\rho^{2}A}\Big) ^{1/2}, \; \; \beta^{\phi}= \dfrac{1}{\alpha}\Big(  \dfrac{2Mr\tilde{a}}{\rho \sqrt{A}}\Big) \,\ .
\end{equation}
$\alpha$ is real if and only if $r>r_{in}$. The keplerian velocity $w_{K}$, in the ZAMO frame, becomes 
\begin{equation} \label{vk}
\hat{v}_{K} = \dfrac{1}{\alpha}\Big[ \sqrt{g_{\phi \phi}} w_{K} -\alpha \beta^{\phi} \Big]
\end{equation}
with $w_{K}$ given by (10). We emphasize that in our numerical computations, we expressed all the quantities in mass unit, so to work with adimensional values. This is equivalent to setting $M = 1$ everywhere .

\section{Energy Extraction} \label{sec3}

\begin{figure}
\centering
\includegraphics[width=0.8\textwidth]{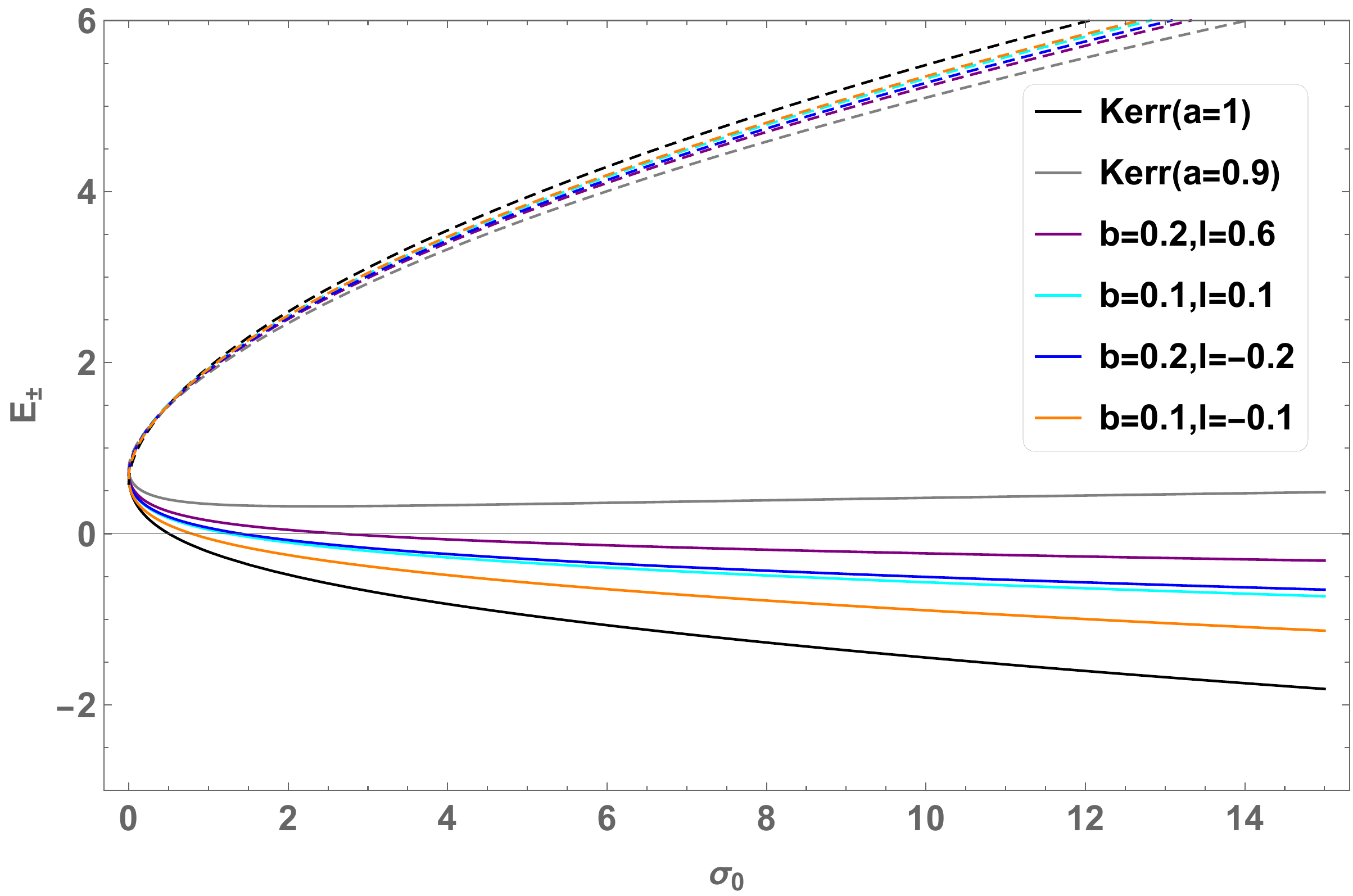}
\caption{Energy at infinity per enthalpy, Eq.~(\ref{eq14}), for accelerating (solid) and decelerating (dashed) plasma for different value of the parameters $\ell$ and $b$. Black and gray curves represent Kerr case, i.e. Lorentz preserving and zero charge $(\ell=0, b=0)$. Adimensional spin values are $a=0.7$ for the purple curve, $a=0.9$ for cyan and gray curves, and $a \rightarrow 1$  in all other cases. Furthermore, we took $\xi=\pi/12$ and imposed $r=r_{isco}$ for each different curve (\textit{model}). The plotted models all have efficiency $\eta > 1$ and maximum extractable energy with respect to Kerr case, i.e. $\eta_{rot} \geq 29\%$. Finally, keep in mind that optimal (but too unrealistic) conditions would be $\xi=0$ and $r=r_{in}$, where $r_{in}$ the outer event horizon. All quantities on the axes are adimensional}
\label{fig:2} 
\end{figure}

Energy extractable by magnetic reconnection deeply depends on the plasma fluid-dynamic and electromagnetic properties. Assuming a one-fluid plasma, the stress-energy tensor is:
\begin{equation} \label{T}
 T^{\mu \nu} = pg^{\mu \nu} + w u^{\mu}u^{\nu} + F\indices{^\mu _\sigma}F\indices{^\nu ^\sigma} - (1/4)g^{\mu \nu}F^{\alpha \beta}F_{\alpha \beta} \,\ ,   
\end{equation}
where $p$, $w$, $u^{\mu}$ and $F^{\mu \nu}$ are plasma pressure, enthalpy density, velocity and electromagnetic tensor, respectively. Neglecting electromagnetic component (assuming a highly efficient transformation of magnetic energy into kinetic energy), the energy density at infinity  \cite{Comisso} is
\begin{equation} \label{einf}
e_{\infty} = \alpha [ (\hat{\gamma}+b^{\phi}\hat{\gamma}\hat{v}^{\phi})w - p/\hat{\gamma} ] \,\ ,
\end{equation}
where $\hat{\gamma} = \hat{u}^{0}$ is the Lorentz factor.  By considering both accelerating and decelerating plasma and dividing by enthalpy, this can be rewritten as \cite{Comisso}:
\begin{equation}\label{eq14}
\begin{split}
E_{\pm}^{\infty} =& \alpha \hat{\gamma}_{K} \Big[
(1+\beta^{\phi} \hat{v}_{K})(1+\sigma_{0})^{1 / 2} \pm \cos \xi(\beta^{\phi}+\hat{v}_{K}) \sigma_{0}^{1 / 2} \\
&-\frac{1}{4} \frac{(1+\sigma_{0})^{1 / 2} \mp \cos \xi \hat{v}_{K} \sigma_{0}^{1 / 2}}{\hat{\gamma_{0}}_{K}^{2}(1+\sigma_{0}-\cos ^{2} \xi \hat{v}_{K}^{2} \sigma_{0})} \Big],
\end{split}
\end{equation}
where $\xi$ is the angle between outflow plasma velocity and the direction $d/d\phi$ at the equator plane, $\gamma_{K}=(1-\hat{v}_{K}^{2})^{-1/2}$ and $\sigma$ is plasma magnetization. An estimate of $\xi$ is given by $\xi = \arctan(v_{r}'/v_{\phi}')$ where prime indicate a calculation in a local rest frame. Since numerical simulations prefer small values~\cite{angleOrientationNumSimulat}, we take as a reference value $\xi=\pi/12$ (note that the best condition for energy extraction would be $\xi \simeq 0$). As in the Penrose process, energy extraction occurs when both the following conditions are satisfied
\begin{equation}\label{eq15}
    E_{-}^{\infty} < 0\; \; \; ,  \; \; \; \; \Delta E_{+}^{\infty}=E_{+}^{\infty}-\Big[ 1- \dfrac{\Gamma}{4(\Gamma-1)} \Big] > 0
\end{equation} 
where we assumed a hot relativistic plasma, for which $w=4p$. Taking a polytropic index $\Gamma=4/3$, one finds the simpler relation $\Delta E_{+}^{\infty}=E_{+}$. An appropriate quantity to define is the energization efficiency $\eta = E_{+}^{\infty}/(E_{+}^{\infty}+E_{-}^{\infty})$. For $\eta > 1$, rotational energy extraction occurs. By using (11) and (12) in (13), different considerations are possible about conditions (14). In Fig. \ref{fig:2}, energy per enthalpy for accelerating ($E_{+}$) and decelerating ($E_{-}$) plasma are shown for different set of parameters $a$, $\ell$ and $b$.
%%%%%%%%%%%%%%%%%%%%%%%%%%%%%%%%%%%%%%%%%%%%%%%
%%%%%%%%%%%%%%%%%%%%%%%%%%%%%%%%%%%%%%%%%%%%%%%%
Although the best X-point distance $r/M$ for energy extraction is $r=r_{in}$, we have chosen to compare the different models at $r=r_{isco}$ since at this distance it is easier to appreciate any differences ($r_{in}\approx 1$ in all cases). We see that Kerr case allows energy extraction ($E_{-}<0$) at lower sigma values, but the deviation from other cases is not significant, even when we consider spin values significantly lower than 1. However, passing from $a\simeq 1$ to $a=0.9$, Kerr becomes unfit to extract energy. On the contrary, bumblebee Kerr-Sen allows the drainage of rotational energy at different spin values. The reason is in the different value of $r_{isco}$ which increases from $\simeq 1$ to $\simeq 2.3$ when $a$ decreases from $1$ to $0.9$, lying beyond the static limit. For the model $(\ell=0.6, b=0.2)$ the ISCO radius is significantly lower, $r_{isco}\simeq 1.45$. Incidentally note that having $r_{isco}<r_{out}$ is not a  necessary  condition to have energy extraction, which strongly depends on proper parameters ($\sigma_{0}$, $\xi$ and $r/M$). However, this condition seems the most natural scenario for magnetic reconnection, as in this case plasma is stably orbiting inside the ergosphere. \\
Another consequence of having a charge is a generalization of irreducible mass (1):
\begin{equation} \label{Mirr}
    M_{irr} = \dfrac{1}{2}\Big[ \tilde{a}^{2} + \Big( 1- \dfrac{b}{2} + \dfrac{1}{2}\sqrt{(b-2)^{2} -4\tilde{a}^2} \Big)^{2} \Big]^{1/2}
\end{equation}
in mass unit. Notice that $E_{max}^{rot}$ increases with increasing $b$, and this means being able to extract energy also at smaller spin values, contrary to Kerr, where $E_{rot}^{max}$ decays quickly even at small decreases of $a$. For example, for a bumblebee-Kerr-Sen black hole with $a=0.7$ and $b=0.2$, the maximum rotational extractable energy is  for $\ell=0.6$ and is equal to $\eta_{rot} \simeq 31 \%$, while the for a classical Kerr black hole with $a=0.9$ is just $\eta_{rot}\simeq 15 \%$. 

\section{Power Extracted Compared to BZ Mechanism} \label{sec4}
\begin{figure}
\centering
\includegraphics[width=1\textwidth]{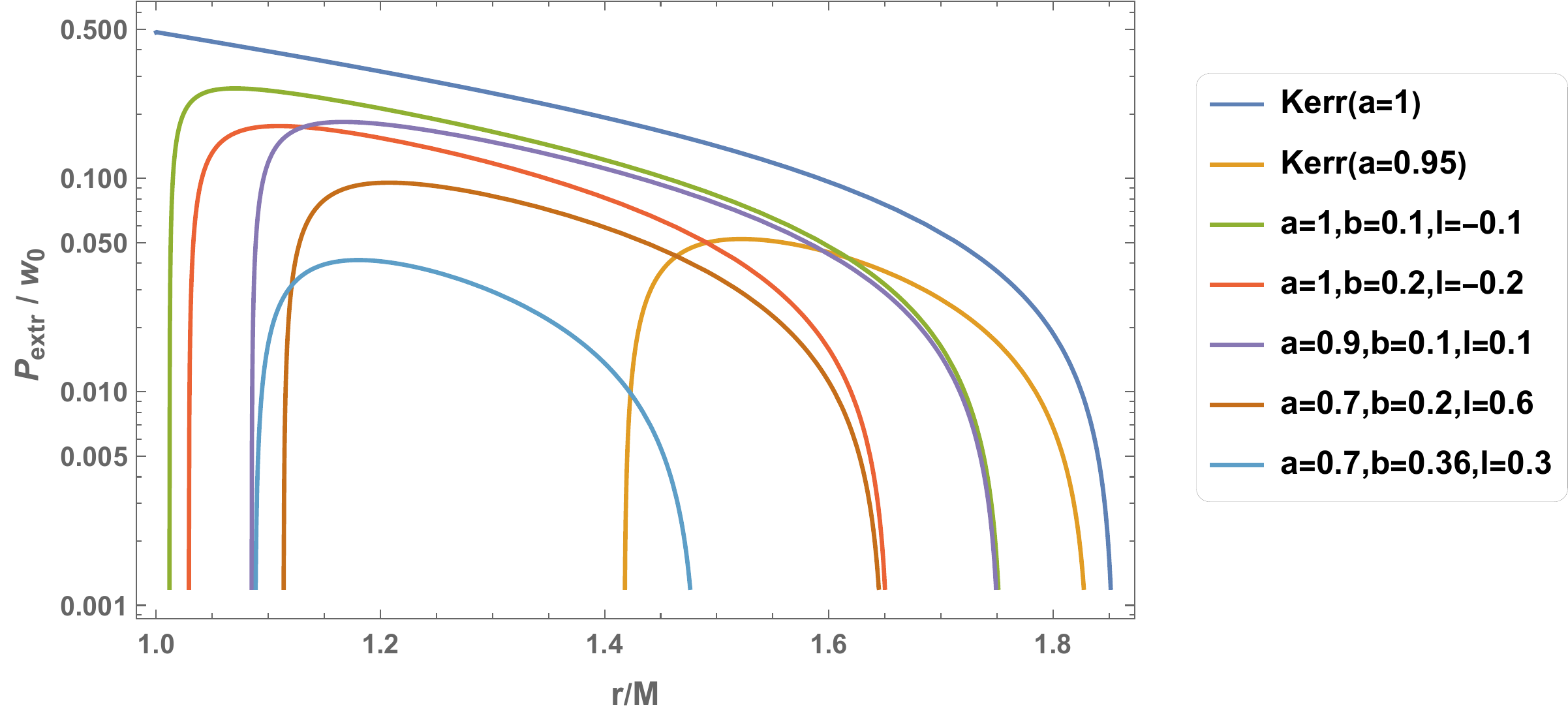}   
\caption{Extracted power via magnetic reconnection per unit enthalpy as function of the X-point position, with $\xi=\pi/12$ and $\sigma=10$, for several set of parameters. The trends do not change much as $\sigma_{0}$ increases. }
\label{fig:3}
\end{figure}

\begin{figure*}[t!!!]
\centering
\vspace{0.cm}
\includegraphics[width=0.9\columnwidth]{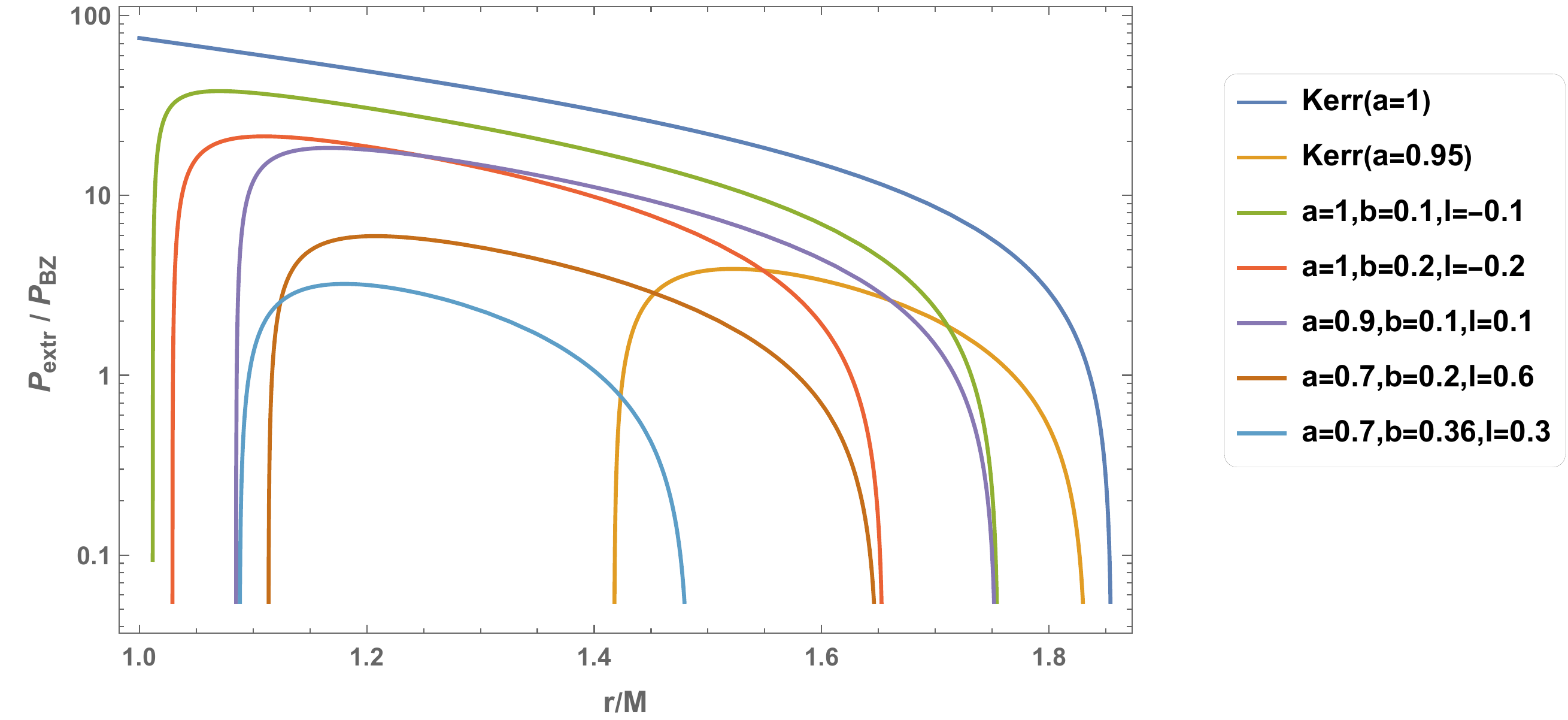}
\hspace{1.cm}
\includegraphics[width=0.9\columnwidth]{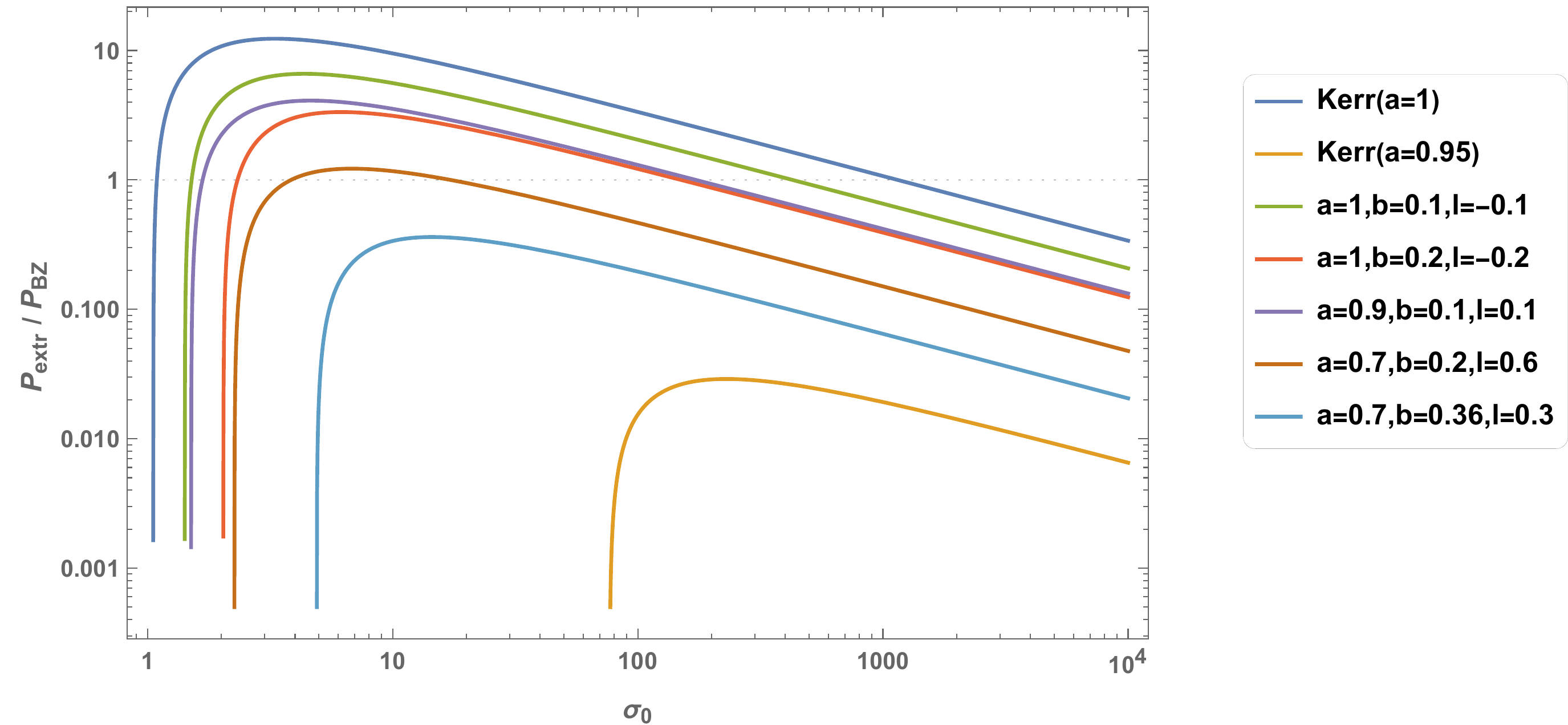}
\caption{(a) Rate between power extracted from magnetic reconnection and the same quantity from BZ mechanism, as function of the X-point distance, with $\xi = \pi/12$ and $\sigma_{0}=10$. (b) The same quantity as function of sigma magnetization, with $\xi = \pi/12$ and $r/M=1.4$. Notice that the choice of the X-point reconnection point $r$ is crucial.}
\label{fig:4}
\end{figure*}

In this section, we compute the rate of energy extraction, $P_{extr}$, in order to evaluate the role of the three parameters $a$, $b$ and $\ell$. Indeed, unlike what \cite{Comisso} and \cite{mohsen} considered, we do not take fast-spinning black holes a priori. To have an approximate relation for $P_{extr}$, we first note that it depends on the amount of plasma with negative energy at infinity in the ergosphere, directed towards  the (outer) event horizon . Because of energy conservation, the "theft" of rotational energy is equal to \cite{Comisso} 
\begin{equation} \label{eq17}
    P_{extr} = -E_{-}^{\infty}w_{0}A_{in}U_{in}
\end{equation}
where $A_{in}$ is the cross-sectional area of the inflowing plasma, which can be estimated with $A_{in}\sim (r_{out}^{2}-r_{ph}^{2})$, assuming spherical shape for the ergosphere; $U_{in}\approx 10^{-1}$ for the collisionless regime and $U_{in}\approx 10^{-2}$ for the collisional one. In order not to get overestimates, we take the first case. In Fig.\ref{fig:3}, we report the extracted power for unit enthalpy, $P_{extr}/w_{0}$, for different set of parameters, including Kerr case, as function of the X-point position, $r/M$.
As it is clear,  extremal Kerr is the most favoured case, but a little decrease in the spin value (just from $a\simeq 1$ to $0.95$) reverses the roles: non-null $\ell$ and $b$ values, allow higher extracted power, even at lower spin values. Again, lower charge values are better,  in line with what was said in Sec. 2. Finally, notice that small charge values allow us to neglect the interaction between the intrinsic magnetic field of the black hole and the external one (from the accretion disk), whose line configuration is essential for the extraction. \\
It is instructive to compare the rate found with the power extracted from the BZ process, which is to date the most popular mechanism for energy extraction from black holes, i.e. \cite{BZ}

\begin{equation} \label{eq18}
    P_{BZ} = \dfrac{k}{16 \pi}\Phi_{BH}^{2}\Omega_{H}^{2}\Big( 1+c_{1}\Omega_{H}^{2}+c_{2}\Omega_{H}^{4}  \Big)
\end{equation}
where $k=0.05$, $c_{1}=1.38$, $c_{2}=-9.2$  are numerical constants. Here, $\Phi_{BH} $ is magnetic flux crossing the BH horizon equal to $2\pi\int_{0}^{\pi}\sqrt{-g} |B^{r}| d\theta$, and $\Omega_{H}$ is the angular velocity of the event horizon,  i.e. $\Omega$ of Fig.\ref{fig:1} (b), evaluated at $r=r_{H}$, which in our notation corresponds to $r_{in}$.  Using  the metric (4) with $a=0$, one gets
\begin{equation*}
    \Phi_{BH}=4 \pi B_{0} \sin{(\xi)}  r_{in}(r_{in}+b) \sqrt{1+l}, \; \; \; \Omega_{H}=\Omega (r=r_{in})=\dfrac{a\sqrt{1+l}}{2 r_{in}},
\end{equation*}
where we used $B_{0}\simeq \sqrt{\sigma_{0}}$. From Eqs.~(\ref{eq17})  and (\ref{eq18}), we can now compute the rate 
\begin{equation} \label{eq19}
    \dfrac{P_{extr}}{P_{BZ}} = \dfrac{-4 E_{-}^{\infty} (r_{out}^{2} - r_{ph}^{2}) U_{in} }{k \pi \sigma_{0} \sin{(\xi)}^{2} (r_{in}+b)^{2}(1+l)^{2}a^{2}\Big(   1 + c_{1}\dfrac{a^{2}(1+l)}{4 r_{in}^{2}} + c_{2}\dfrac{a^{4}(1+l)^{2}}{16 r_{in}^{4}}  \Big)  }
\end{equation}
where $E_{-}^{\infty}$ is given by Eq.~(\ref{eq14}). In Fig. \ref{fig:4} we plotted Eq.~(\ref{eq18}) as function of the X-point distance or plasma magnetization.

As already emerged about Fig. \ref{fig:3}, extremal Kerr has the maximum extracted power. However, modification arising from $\ell,b \not= 0$ deeply affect the ratio $P_{extr}/P_{BZ}$, privileging magnetic reconnection. Indeed, $P_{extr}/P_{BZ}>1$ for several set of parameters. In particular, the model $(a=0.7,b=0.2,\ell=0.6)$  offers performance similar (and in some points superior) to the Kerr case $a=0.95$, despite having a much lower spinning. Note also that the latter case has a smaller range of possible X-points, a consequence of a larger $r_{in}$ as explained in Sec. 2, and that in this case the maximum efficiency would be $\eta_{rot}\simeq 19\%$, far below the corresponding value for the bumblebee Kerr-Sen case, $\eta \simeq 31\%$. This indicates that more massive black holes would be needed to extract the same amount of energy. We underline the choice of the X-point position when we analyze the trend as a function of $\sigma_{0}$, since each model has a specif range of extraction.  In general, it appears clear that magnetic reconnection in purely Kerr black holes needs higher plasma magnetization as $a$ decreases, indicating in this case it can happen only in highly magnetized systems. On the contrary, the bumblebee-Kerr-Sen contest turns out to be not only highly energetic but also superior to BZ and at lower $\sigma_{0}$. The order of magnitude of the BZ process is \cite{Lee_2000}
\begin{equation*}
    P_{BZ} \sim 6.7 \times 10^{50} \Big( \dfrac{B}{10^{15} G}\Big)^{2} \Big( \dfrac{M}{M_{\odot}} \Big)^{2} \; erg / s 
\end{equation*}
where $B$ in the magnetic field. One finds that for a not-supermassive black hole of $10 M_{\odot}$, and a magnetic field of $B\sim 10^{14} G$,  typical values are $P_{BZ}\sim 10^{51}$ $erg/s$. From Fig. \ref{fig:4} (b), this means that magnetic reconnection in this case would extract up to $\sim 10^{52}$ $erg/s$, provided a strong enough magnetic field. This would explain GRBs from MR in not-supermassive black holes not only for extreme Kerr black holes, but also for not-extreme charged black holes, since MR in Kerr case lose position (respect to BZ) very quickly when spinning decreases. 
\section{Analysis on the parameter space} \label{sec5}
\begin{figure*}[t!!!]
\centering
\vspace{0.cm}
\includegraphics[width=0.46\columnwidth]{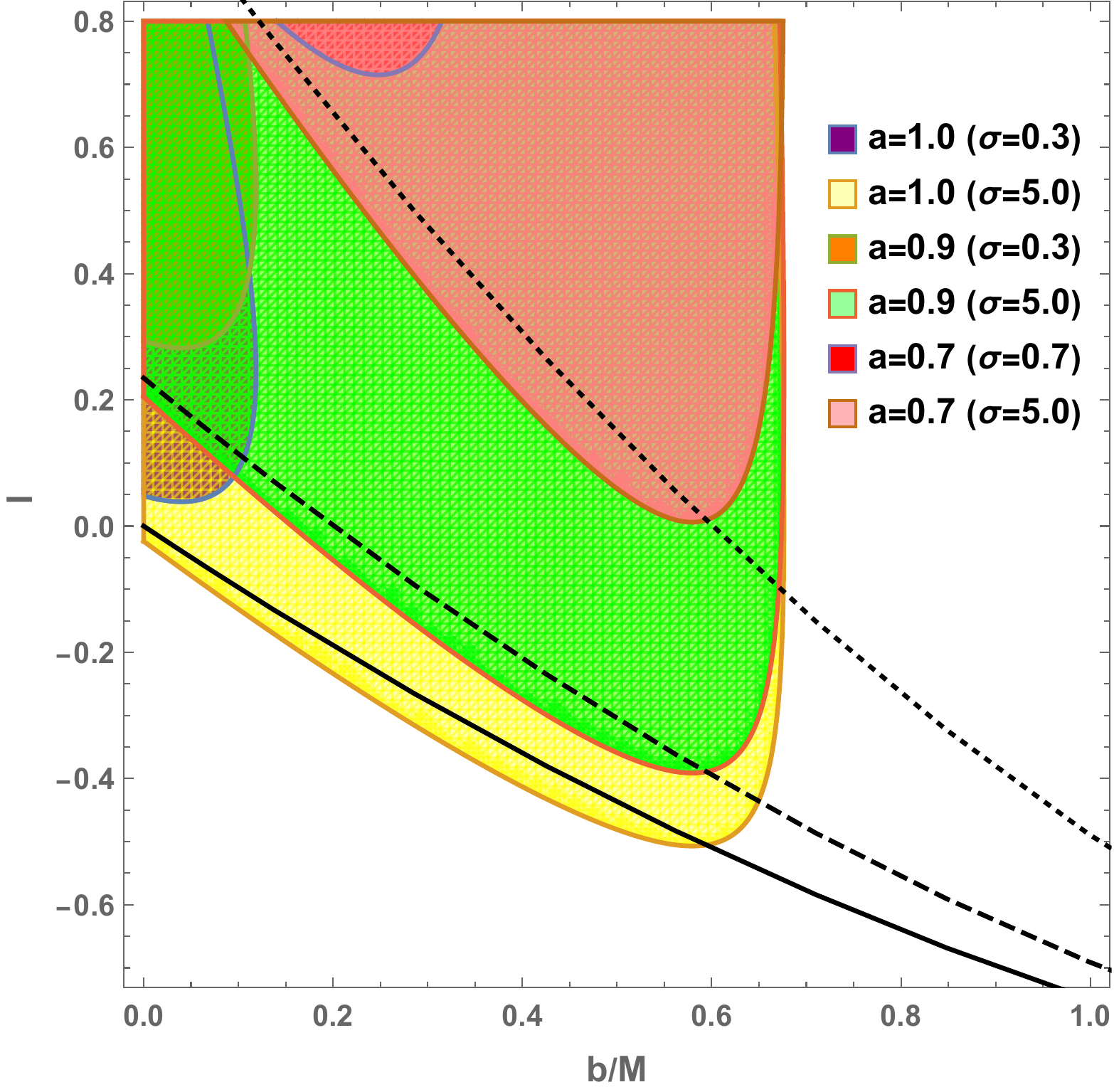}
\hspace{1.cm}
\includegraphics[width=0.453\columnwidth]{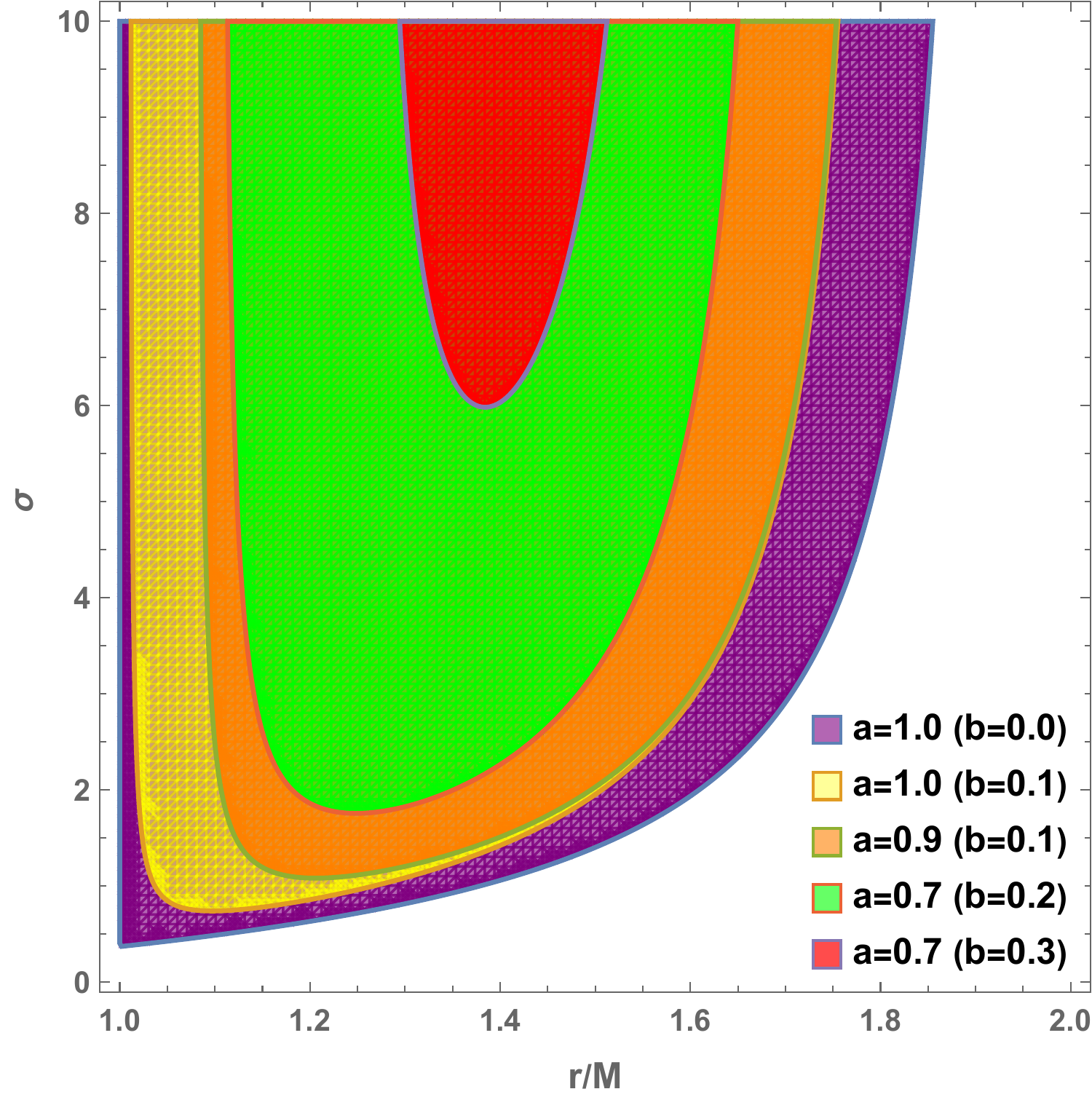}
\vspace{0.5cm}
\includegraphics[width=0.7\columnwidth]{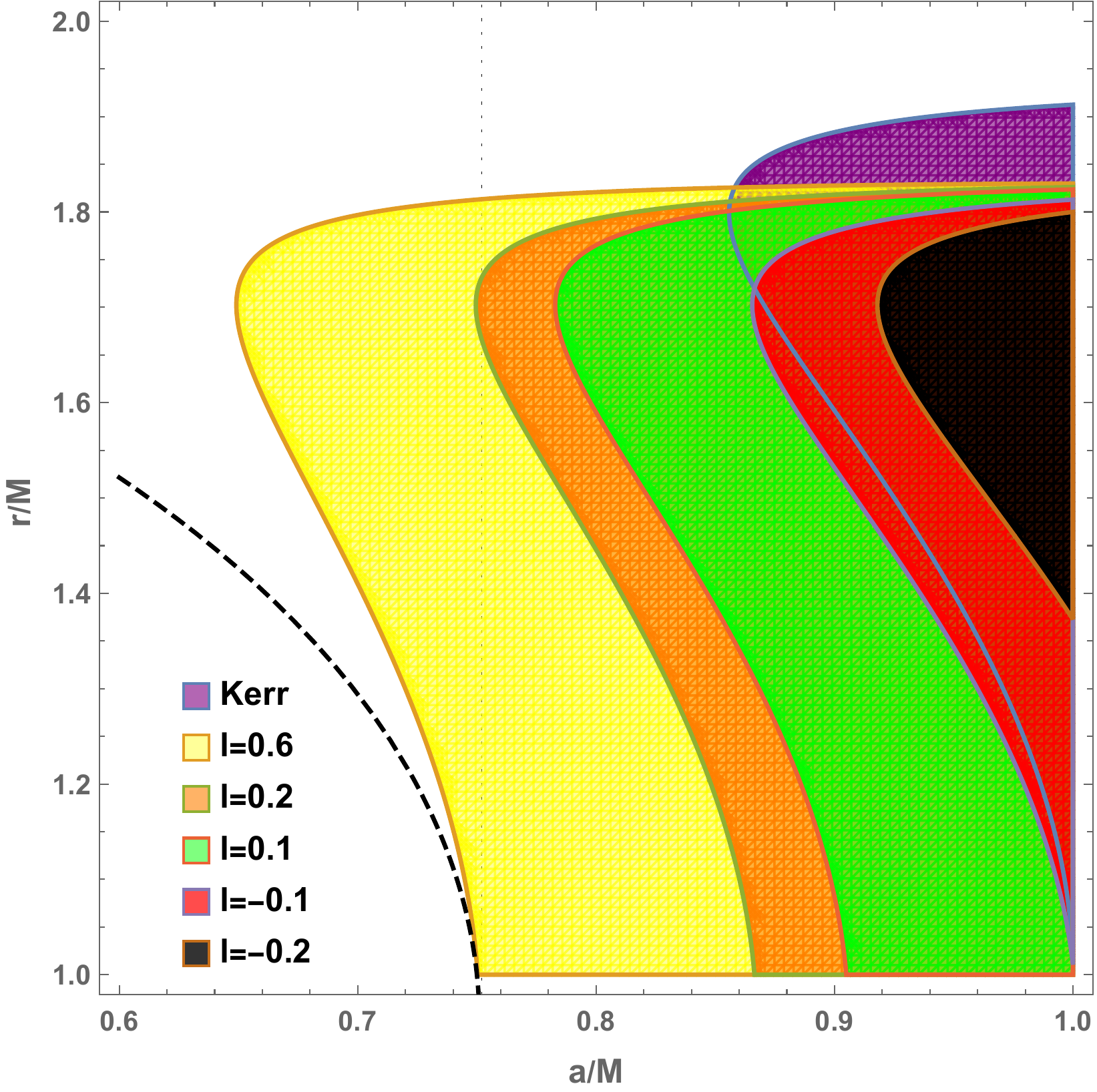}
\caption{(a) Parameter space with $\ell$ and $b$ for different values of $a$ and $\sigma$ and with $r=1.2$ and $\xi=\pi/12$ fixed. Solid, dashed and dotted curves represent existence condition of an event horizon for $a=1$, $a=0.9$ and $a=0.7$ respectively. Only portions of regions below these curves have an event horizon, otherwise naked singularities have to be considered. As $a$
decreases, it is necessary to increase $\sigma_{0}$ to have energy extraction; while raising the X-point radius $r/M$ ,instead, shrinks regions
towards positive $\ell$ and smaller b values.   (b) Parameter space $\sigma_{0}$ vs $r/M$ for different options of $a, b$ and $\ell$ and with $\xi=\pi/12$ fixed. In this figure we plotted models with $r_{ISCO} < r_{out}$, which is the most natural scenario for magnetic reconnection to occur. The $\ell$ values are the same of Fig. \ref{fig:2} ($\ell=0.3$ for the case $b=0.3$).  (c) Parameter space $r/M$ vs $a/M$, with  $\sigma_{0}=100$, $b=0.1$ and $\xi=\pi/12$ fixed. Notice that adimensional values $a < 0.9$ are allowed, contrary to the Kerr case. The dashed  curve represents  the event horizon limit, $r_{H}=r_{in}$, for the case $l=0.6$ while the vertical dotted line is the upper bound on $a$ for this case: to avoid naked singularities, only portions on the left of this line are allowed. A very similar behavior applies to the other cases. }
\label{fig:5}
\end{figure*}

As has already been mentioned, the choice of the parameters $a$, $b$ and $\ell$ which define each \textit{model} is decisive. If we add to these the parameters of energy production, i.e.  the orientation angle $\xi$, the plasma magnetization $\sigma_{0}$ and the X-point position $r/M$, an analytical analysis is useless, and only a parametric approach is possible.  In particular, in Fig. \ref{fig:5} different parametric approaches are shown.

In sub-figure (a), we take as parameters $\ell$ and $b$ and show the region in which condition(\ref{eq15}) is true, for various values of $a$ and $\sigma_{0}$, and with $\xi$ and $r/M$ fixed. In the range $\sigma_{0}<1/3$ there is no extraction for a classical Kerr BH (according to \cite{Comisso}), but turning on the extra parameters, a  non-zero region appears even for $\sigma_{0}=0.3<1/3$ and also for $a<1$  \footnote{We found that even smaller $\sigma_{0}$ values are allowed (like $\sigma_{0}=0.2$), although the corresponding region in which condition (\ref{eq15}) is satisfied would be very tight. Contrary to \cite{Comisso}, we use do not use here the absolutely best conditions, using which we should have take $r=r_{in}$ for each point of the parameter space (remember that $r_{in}$ depends on $\ell$ and $b$) and $a \rightarrow 1$ everywhere. This make generalized Kerr solutions even more promising. } (see orange region) . This means that $\sigma_{0}=1/3$ is not an upper limit for this extended Kerr solution. But an important clarification must be made: only the regions that lie below the solid, dashed and dotted curves (for $a=1$, $a=0.9$ and $a=0.7$, respectively) have an event horizon (see Sec. 2), otherwise one has to consider naked singularities.  Notice also that, as we expected,  as $a$ decreases, it is necessary to increase $\sigma_{0}$ to have energy extraction; for $a = 0.7$, the minimum $\sigma_{0}$ value is $\sim 0.7$. Raising $r/M$, instead,  shrinks regions towards positive $\ell$ and smaller $b$ values.   Using $r/M=1$ (the smallest possible value) and fixing $\sigma_{0}$, a maximum allowable $b$ value becomes visible. For example, with $\sigma_{0}=5$ it  seems to be $b \leq 0.9$; while increasing $\sigma_{0}$ to infinity, the limit changes very little ($b \leq 0.95$), never reaching $b=1$.  This is not an intrinsic constraint on the charge, but only a condition for having energy extraction.  \\
In sub-figure (b), the plasma magnetization and the X-point radius are considered parameters, while we used different options for $a, b$ and $\ell$. In this figure we  exhibited models for which $r_{ISCO} < r_{out}$, like in Fig. \ref{fig:2}. As already mentioned, lower spin values are associated with smaller extraction regions, towards higher plasma magnetization. Again, low charge values are favoured.\\
Finally, in sub-figure (c) the space parameter shows the X-point position and the (adimensional) spin value. We fixed $\sigma_{0}=100$ and a little charge, $b=0.1$, in order to light up the effect of the Lorentz breaking parameter. It soon becomes clear that not-extreme spinning black holes are allowed, and the bumblebee Kerr-Sen solution deeply extends Kerr results. Higher values of $\ell$ are now favoured but one carefully has to take extraction regions, since, as mentioned, naked singularities are lurking. The condition of the existence of an event horizon is translated into an upper bound on the spin value for each value of $\ell$. 
\section{Magnetic reconnection with dark energy Quintessence} \label{sec6}
Astrophysical BHs are not isolated from matter. Still today,  it is not yet clear what kind of matter dominates the region around the BHs, and we expect a certain variety of cases. The effect of dust, radiation and dark matter on superradiance (assumed as energy extraction way) was only recently studied \cite{superradianceDE}. In this section, we investigate how this type of energy-matter around a (Kerr) black hole can affect energy extraction through magnetic reconnection. \\
An extended Einstein solution which recently is attracting a lot of interest,  is quintessence \cite{quintes1,quintes2}. Quintessence is a scalar field with negative pressure and equation of state (EoS) $p_{de}=w_{de}\rho_{de}$, where $p_{de}$ is the pressure, $\rho_{de}$ the energy density and $-1<w_{de}<-1/3$ the state parameter of the dark energy component. The case $w<-1$ corresponds to the so-called phantom energy, while $w=-1$ corresponds to cosmological constant.

\subsection{Rotational Kiselev black holes}
\begin{figure*}[t!!!]
\centering
\vspace{0.cm}
\includegraphics[width=0.8\columnwidth]{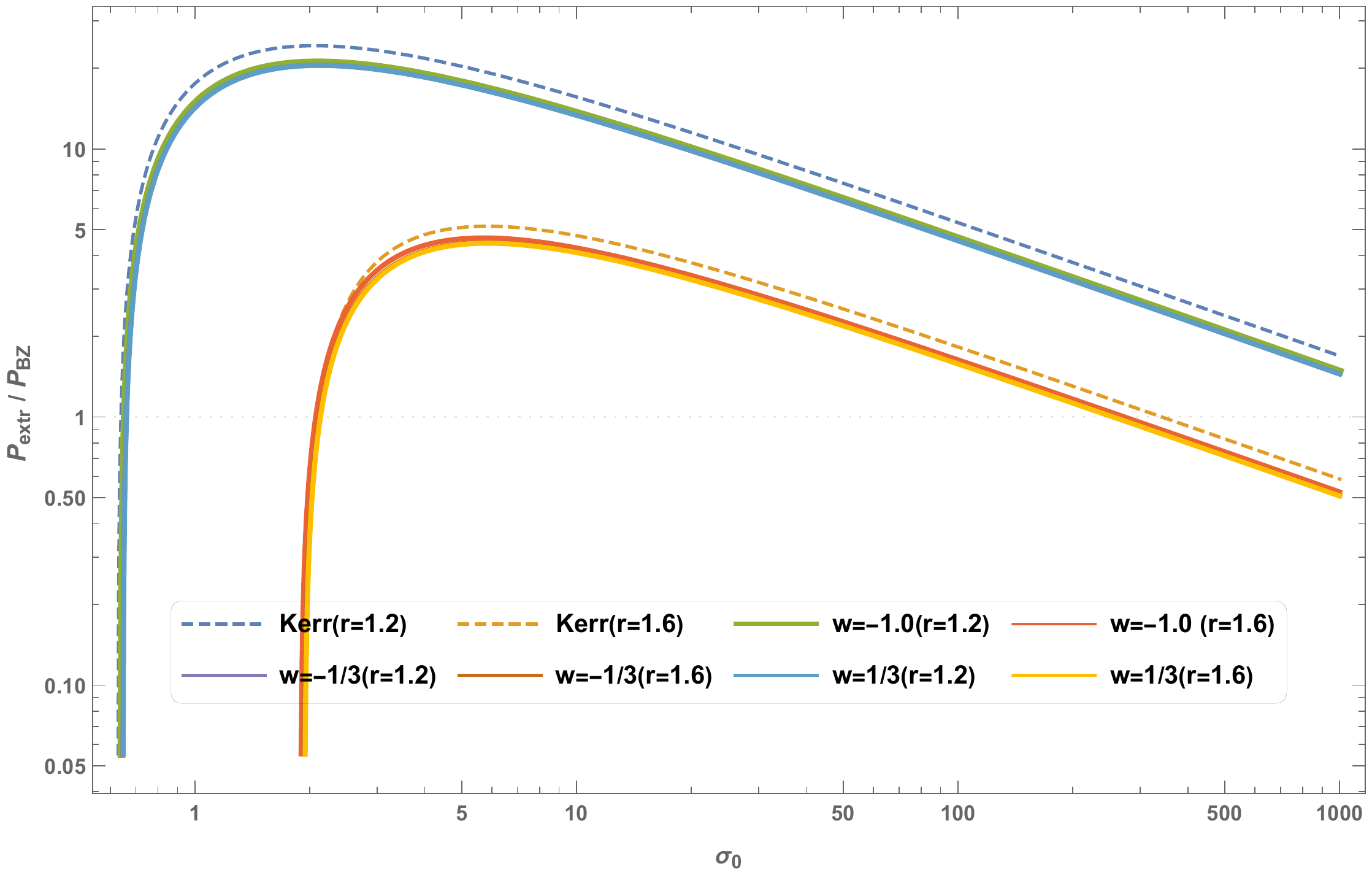}
\hspace{1.cm}
\includegraphics[width=0.7\columnwidth]{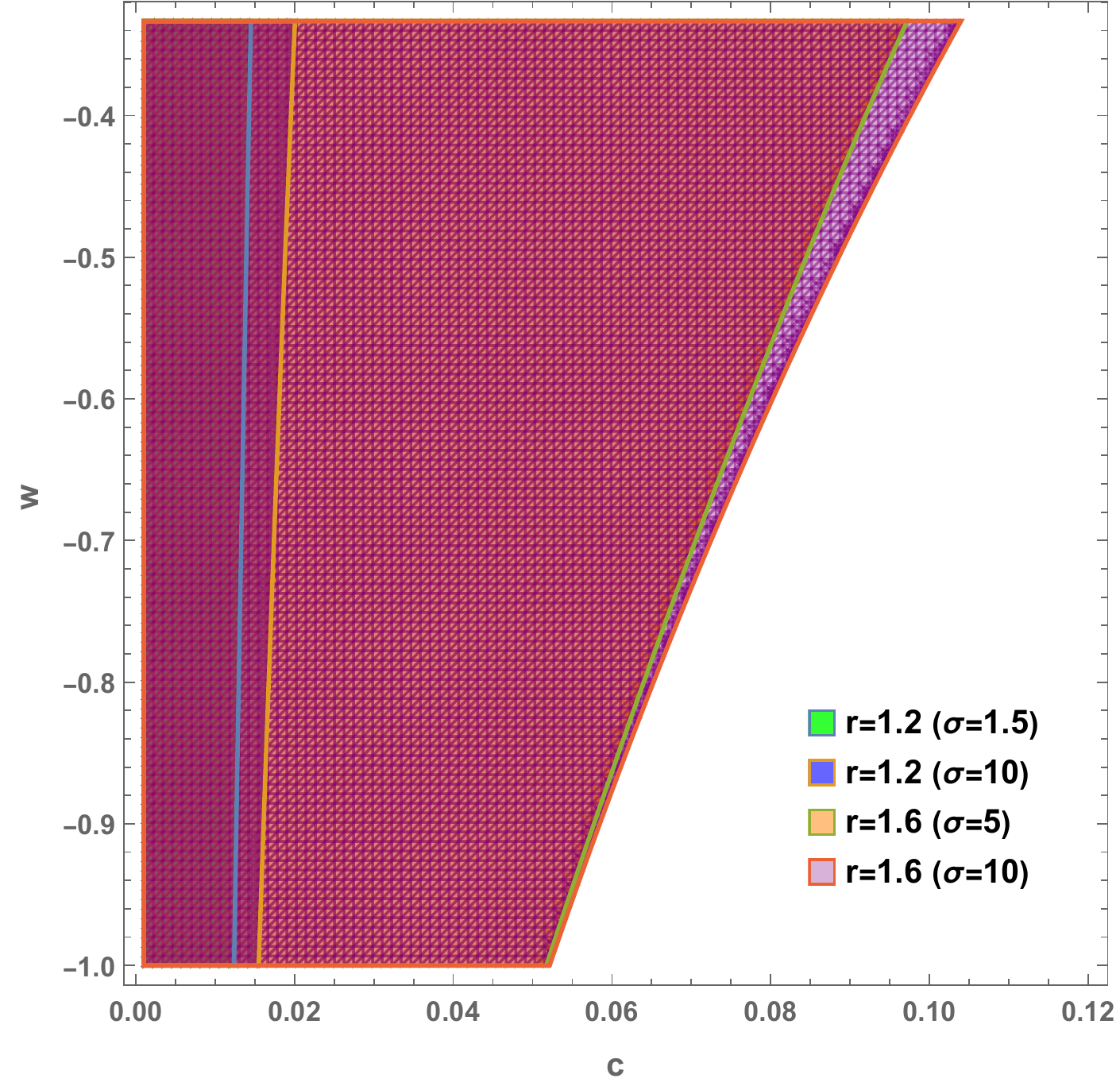}
\caption{(a) Rate between power extracted from magnetic reconnection, $P_{extr}$, and the same quantity from BZ mechanism, $P_{BZ}$, as a function of the X-point distance, with $\zeta = \pi/12$ and the optimal conditions $a\rightarrow 1$ and $c=1/1000$. Notice the very similar behaviour between a Kiselev BH and the Kerr case, and also for different values of $w$ (only some shown). Since here we assumed $a \rightarrow 1$, we computed $r_{+}$ from $\Delta=0$, rather than from Eq.~(\ref{eq21}).  (b) Parametric space $w$ vs $c$ for which condition (14) is true. Different options for $\sigma_{0}$ and $r$ are shown. Notice that increasing the distance of the latter or increasing the magnetization increases the range of $c$ values for which there is extraction, while no bound on $w$ is evident. }
\label{fig:6}
\end{figure*}

The solution of Einstein's field equation for a Schwarzchild black hole surrounded by quintessence has been obtained in Ref.~\cite{Kiselev:2003,Ghosh}. Even if in this works only quintessence is considered, the Kiselev solution contemplates any type of energy-matter, once a state parameter has been established. Indeed, a rotational Kiselev black hole looks like \cite{Toshmatov:2015npp}

\begin{equation}
\begin{aligned}{}
d s^{2}=-\Big(1 & -\frac{2 M r+ c r^{1-3 \omega}}{\Sigma^{2}}\Big) d t^{2}  +\frac{\Sigma^{2}}{\Delta} d r^{2}-\frac{2 a \sin ^{2} \theta\left(2 M r+ c r^{1-3 \omega}\right)}{\Sigma^{2}} d \phi d t \\
&+\Sigma^{2} d \theta^{2}
    +\sin ^{2} \theta\left(r^{2}+a^{2}+a^{2} \sin ^{2} \theta \frac{2 M r+ c r^{1-3 \omega}}{\Sigma^{2}}\right) d \phi^{2}
    \label{Eq:quintessence}
\end{aligned}
\end{equation}

where we defined
\begin{equation*}
 \Delta=r^{2} -2Mr +a^{2}-cr^{1-3w}, \; \; \;  \Sigma^{2}=r^{2}+a^{2}\cos{\theta}^{2} . 
\end{equation*}
As before, $M$ is the mass of the black hole and $a$ is the spin parameter. Moreover, $c$ is the strength parameter and $w$ defines the EoS, $p=w\rho$. Eq.~(\ref{Eq:quintessence}) is the rotational symmetry solution for a black hole wrapped in any kind of energy-matter definable by the EoS. In general, for dark energy, we would expect $w<0$; quintessence, as said, would satisfy a specific range, i.e. $-1<w<-1/3$. We also investigate dust ($w=0$) and radiation ($w=1/3$), as well as cosmological constant ($w=-1$) and the so-called $R_{h}$ universe \footnote{As the $\Lambda$CDM model, also  this Einstein cosmological solution is Friedman-Robertson-Walker-based, but although the two theories have much in common,  it has precisely the additional constraint $p=-(1/3)\rho$, where we point out that now $p= p_{m}+p_{r}+p_{de}$ and $\rho= \rho_{m}+\rho_{r}+\rho_{de}$. In other words, remembering that $w=0$ for dust and $w=1/3$ for radiation, in this model $w=\rho/3+w_{de}\rho_{de}=-1/3$, i.e. it is the \textit{total} state parameter to be fixed (at all times), allowing $w_{de}$ to be different from $-1$ as usually imposed. In the following, we simply use $w$ to indicate the state parameter, with the caveat that when $w=-1/3$ we are referring to the total equation of state, and not only to the dark energy component. Indeed, being $w=-1/3<0$, it could refer to pure dark energy as well. } ($w=-1/3$) \cite{Melia:2014vva}. \\
The number of horizons depends on the value of $w$. For $-1\leq w <-1/3$, $\Delta=0$ has three positive solutions, corresponding to a Cauchy horizon, an event horizon and a cosmological horizon. For $w=\pm1/3$ and $w=0$, the cosmological horizon disappears and only two horizons exists. \\
Since in our paper we consider $c\ll 1$, it is possible to use perturbation method to calculate horizon radius, as a perturbation of the Kerr horizon. At the first order of the strength parameter $c$, the outer horizon  becomes \cite{Xu}
\begin{equation} \label{eq21}
    r_{+} \simeq M + \sqrt{M^{2}-a^{2}}+\dfrac{c(M+\sqrt{M^{2}-a^{2}})^{1-3w}}{2\sqrt{M^{2}-a^{2}}}
\end{equation}
which is valid for $a\not= 1$.
In order to find an upper limit on the strength parameter $c$, we can evaluate the maximum spin value. From $\Delta=0$, we solve in function of $c$ the maximum condition
\begin{equation} \label{stre}
    \dfrac{\partial}{\partial r} \Big( c r^{1-3w} +2 M r - r^{2} \Big) = 0.
\end{equation}
For example, when $w=-1$, the solution is $c=(2r-2)/(4r^{3})$, whose  maximum value is $c=2/27$ at $r=3/2$, hence, in this case, $c\leq 2/27$. Similarly, when $w=-2/3$ or $w=-1/3$, one finds $c\leq 1/6$ and $c\leq 1$, respectively. Finally, when $w=0$ or $w=1/3$ no useful limits turns up. We emphasize that this conditions on $c$ are necessary but non sufficient to have (real) event horizons. Furthermore, they do not insure real ISCO radius. Indeed, when $w=-1$, $r_{ISCO }>0$ only for $c\leq 1/100$.  Hence, this second constraint becomes the effective one.  As before, we looked for the most likely scenario for magnetic reconnection inside the ergosphere, i.e. $r_{ISCO}< r_{out}$, where $r_{out}$ is the static limit. To give an order of magnitude, this happens only for $c\leq 1/1000$ when $w=-1$; for $c \leq 1/100$ when $w=-2/3$; for $c \leq 1/10$ when $w \in \{-1/3, 0, 1/3\}$. Besides, only highly spinning black holes are admitted this time, i.e. $a>0.9$. Since $M_{irr}$ is very sensitive to $c$, lower $c$ values are associated with greater removable rotational energy \footnote{The drag effect, instead, is higher at higher $|w|$ values, and, when $w$ is fixed, is stronger at higher $c$ values. When $c\ll 1$, the dependence on $c$ is linear.}; in the following, we choose $c=1/1000$ when more options are possible. Interestingly, for extremal ($a=1$) black holes, the values of $r_{ph}$, $r_{ISCO}$ and also $r_{in}$ (event horizon) and $r_{out}$, become more and more equal as $c$ decreases,  becoming practically indistinguishable when $c = 1/1000$, regardless of the $w$ value. This is certainly due to the shape of the potential (23). From these first considerations, it is immediately found that, whatever $w$ is, the maximum efficiency, $\eta_{rot}$, never exceeds that of Kerr (see Sec. 1). We point out here that having grater value of the strength parameter could change the  plasma stress-energy tensor $T_{\mu \nu}$ (Sec. 3), depending on the type of coupling between electromagnetic field and dark energy. As long as $c\ll 1$, then the hydrodynamic component of the plasma-fluid will be dominant and no correction will be needed.  \\ 
Then, using (20), one finds for the horizon angular velocity, $\Omega_{H}$, and  the irreducible mass, $M_{irr}$, the following results 
\begin{equation} \label{ome}
    \Omega_{H} = \lim_{r\rightarrow r_{+}}\Big(- \dfrac{g_{t\phi}}{g\phi\phi}\Big) = \dfrac{a}{r^{2}_{+}+a^{2}}, \; \; \; M_{irr}=\dfrac{1}{2}\sqrt{r_{+}^{2}+a^{2}}
\end{equation}
Notice that at the first order $\Omega_{H}$ does not depend on $w$ and $c$. The Keplerian angular velocity, instead, has a more elaborate expression, i.e.

\begin{equation} \label{ww}
    w_{K}=\frac{a \left(c+2 r^{3 w}+3 c w\right)-\sqrt{2}  r^{2+3 w} \sqrt{ r^{(-1-3 w)}\left(c+2 r^{3 w}+3 cw\right)}}{-2 r^{3+3 w}+a^{2} \left(c+2 r^{3 w}+3 c w\right)}
    \end{equation}
which is equal to the Kerr keplerian velocity when $c=0$.
The effective radial potential looks like
\begin{equation} \label{pote}
\begin{aligned}{}
    V_{eff} = r^{-3(1+w)}\Big[r^{3+3 w}+a^{2}\Big(c+r^{3 w}&(2+r)\Big)-2 a S\left(c+2 r^{3 w}\right)  \\ &+S^{2}\Big(c +2 r^{3 w} -r^{1+3 w}\Big)\Big] E^{2}
    \end{aligned}
\end{equation}
where we remember that $S=L/E$ is the impact parameter and $E$ is the energy of the (massive or not) rotating particle. We point out that, as for the photon orbit of bumblebee Kerr-Sen black hole,  the energy is factored out and does not contribute to the calculation of the photon orbit, while it is an unknown for the ISCO radius computation, together with the angular momentum parameter, $S$, and the radius, $r/M$. \\
Magnetic reconnection drains rotational energy if the decelerating plasma energy is negative, i.e. $E_{-}<0$, where $E_{-}$ is given by (13). Then, to make a comparison with the BZ process, we need the magnetic flux which crosses the event horizon, which is equal to $\Phi_{BH}=4 \pi B_{0} \sin{\xi} r^{2}_{+}$, where, as for the Kerr-Sen solution, we fixed $\xi=\pi/12$ (see Sec. 3). Using (17) and (22) and passing to mass unit quantities, we numerically simulate the rate $P_{extr}/P_{BZ}$, where $P_{extr}$ is the power extracted via magnetic reconnection into the ergosphere. We plotted it as function of plasma magnetization, $\sigma_{0}$, in Fig. \ref{fig:6} (a). It turns out quite eloquent that a Kiselev rotational black hole is practically indistinguishable from a Kerr one, and this regardless of the value of $w$, i.e. the type of matter surrounding the black hole. 

Note that, even in this case, the rate is very much in favour of magnetic reconnection with respect to the Blandford-Znajek mechanism. In Fig. \ref{fig:6} (b), a parametric-space analysis was made, for different $\sigma_{0}$ values and X-point positions.  Increasing the distance of the latter or increasing $\sigma_{0}$ increases the range of $c$ values for which there is extraction, while no bound on $w$ is evident.

\section{Conclusion} 
\label{sec7}

High energy astrophysical events impose a tough challenge in finding the physical mechanisms capable of generating such high energies. The most accepted one, the Blandford-Znajek (BZ) mechanism, has recently been overtaken by a new way of extracting rotational energy directly from the black hole \cite{Comisso}. It predicts a phenomenon of magnetic reconnection  inside the ergosphere of rotational BHs. An important signature of this new energy extraction way is its transient and/or intermittent nature, contrary to the continuous nature of the BZ process. The reason for
this bursty behaviour is the time it takes to accumulate magnetic energy, the
storing of which requires appropriate dynamics of the configuration of the
magnetic field lines. This feature reinforces the idea of its role in relativistic jets. For large magnetization values, $\sigma \gg 1 $, they found that the asymptotic negative energy per enthalpy of the decelerated plasma goes like $ e_{-}^{\infty} \simeq - \sqrt{\sigma/3}$, while the accelerated plasma, which escapes to infinity in the opposite direction to the black hole taking away rotational energy at the expense of the latter, behaves like $ e_{+}^{\infty} \simeq  \sqrt{3\sigma}$. \\
In this paper, we evaluated if this extraction way could be achievable in an extended Kerr solution, exactly a Kerr-Sen BH (which is a solution of heterotic string theory) with a bumblebee background. Heterotic string theory is one of the primary candidates to describe quantum gravity, with some relevant differences from GR, which make it visible in crucial phenomenological aspects, such as the shadow of a black hole (see e.g. \cite{Xavier_2020} and \cite{TestKerrSen}). Although it has been shown that a Kerr-Sen black hole has a larger shadow than its general relativity analogue (Kerr-Neumann BH), this effect, already very small, it is negligible when the charge is low. Therefore, we expect that almost all the conclusions we have reached are also valid for a classical charged version, i.e. a Kerr-Neumann black hole, especially when rotation is not extreme. Notice that a proper Kerr-Sen metric is obtained by setting $\ell=0$.    \\  
Contrary to Kerr, the two additional parameters, the charge $b$ and the Lorentz breaking one $l$, make the analysis much more complicated, starting from the computation of the innermost stable circular orbit (ISCO), for which only a numerical evaluation is possible. 
The orbit at $r=r_{isco}$ has crucial importance in magnetic reconnection, and represent the inner boundary of a possible accretion disk, which also is thought to have a role in generating GRBs. We looked for the best cases in which magnetic reconnection occurs and the condition $r_{ISCO}<r_{out}$ is realized, where $r_{out}$ is the external boundary (static limit) of the ergosphere. While 
passing from $a \simeq 1$ to $a = 0.9$, Kerr case becomes quickly unfit to extract energy, the bumblebee Kerr-Sen case allows strong drainage of rotational energy also at different
spin values. The reason is in the different value of $r_{ISCO}$ which increases from
$\simeq 1$ to $\simeq 2.3$ when $a$ decreases from 1 to 0.9, lying beyond the static limit. For
the model $(l = 0.6, b = 0.2)$ the ISCO radius is significantly less, $r_{ISCO} \simeq 1.45$. Even if having $r_{ISCO} < r_{out}$ is not a necessary condition to have
energy extraction (the innermost limit is $r_{ph}$), which strongly depends on proper parameters (plasma magnetization $\sigma_{0}$, orientation angle $\xi$, and the X-point position
$r/M$), this condition seems the most natural scenario for magnetic
reconnection, as in this case plasma is stably orbiting inside ergosphere. Interestingly, given
a value of the charge $b$ (with $b < 0.4$), the smallest value of $\ell$ in order to have
$r_{ISCO} < r_{out}$ is $\ell = -b$. Besides, magnetic reconnection in
purely Kerr black holes needs higher plasma magnetization as $a$ decreases, indicating that in this case it can happen only in highly magnetized systems. On the
contrary, bumblebee-Kerr-Sen contest turns out to be not only highly energetic
but also superior to BZ and at noticeably lower magnetization. The reason is that $r_{isco}$ decreases as a function of $b$ much faster than $r_{out}$ increases as $b$ increases. The other side of the coin is frame-dragging, $\Omega=d\phi/dt$, reduction as spinning decreases. However, while Kerr remains the best case,  in the bumblebee-Kerr-Sen models the decrease is negligible if the values of $\ell$ are chosen appropriately. We also found that the maximum extractable rotational energy $E_{rot}^{max}$
increases with increasing charge $b$, and this means being
able to extract energy also at smaller spin values (like $a=0.7$), contrary to Kerr, where $E_{rot}^{max}$
decays quickly even at small decreases of $a$. From a parametric space analysis, we noted that $\sigma_{0}=1/3$ is not an upper limit for this extended Kerr solution, as it happens for pure Kerr. In general, we found that little charge and positive $\ell$ values are the best for energy extraction, but one carefully has to choose parameters, since good values for energy extraction could have naked singularities: the condition of existence of an event horizon is translated
into an upper bound on the spin value for each value of $\ell$. In the final part of the paper, we apply the same formalism to a Kiselev rotational black hole, another extended Kerr solution, in which the black hole is wrapped by energy-matter defined by an equation of state such as $p=w\rho$. We considered dark energy options, i.e. negative state parameters, and also dust ($w=0$) and radiation $(w=1/3)$. In particular, the value $w=-1/3$ could refer to the so-called $R_{h} =c t$ model, a recent and elegant alternative to the $\Lambda$CDM model, of which it could be a valid extension. Interestingly, we found that this extended solution is practically indistinguishable from Kerr one when the strength parameter is $c=1/1000$, and hence is a good way of extracting energy. Larger values of $c$, on the other hand, hinder the extraction of energy, giving always lower results than the Kerr case, and this is regardless of whether the black hole is surrounded, be it dark energy or ordinary  (dust or radiation). Finally, no bounds on $w$ are possible with a parametric analysis: the magnetic reconnection mechanism seems to be insensitive to the surroundings of the black hole. Future jobs could mainly concern two aspects: (1) investigate if the intrinsic magnetic field of a spinning charged black hole could affect energy extraction by changing external magnetic configuration; (2) put into account a possible coupling between dark energy and plasma electromagnetic field such as to be comparable to hydrodynamic energy in the stress-energy tensor. \\
In conclusion, we found that a pure extremal ($a=1$) Kerr solution is the best case for rotational energy extraction via magnetic reconnection, but it is indistinguishable from a Kerr black hole surrounded by energy-matter when the strength parameter is $c=1/1000$ (when $c=0$ one obtains exactly Kerr). Furthermore, when  $a<1$, pure Kerr loses ground very very quickly, and the presence of even a small charge ($b \simeq 0.1$) strongly improves energy extraction, providing a valid alternative to the BZ mechanism, which remains a good channel only for high plasma magnetization values.    

\begin{acknowledgments}
The work of A.C., G.L. and L.M. is supported by the Italian Istituto Nazionale di Fisica Nucleare (INFN) through the ``QGSKY'' project and by Ministero dell'Istruzione, Universit\`a e Ricerca (MIUR).
 \end{acknowledgments}
 
\bibliographystyle{JHEP.bst}
\bibliography{biblio2.bib}

\end{document}